\documentclass[aps,pre,reprint,superscriptaddress]{revtex4-2}

\usepackage{amsmath,amssymb,bm}
\usepackage{graphicx}
\usepackage[colorlinks=true,citecolor=blue,linkcolor=blue]{hyperref}

\usepackage{physics}
\usepackage{tikz}
\usetikzlibrary{arrows.meta,angles,quotes,decorations.markings,calc}

\newcommand{\pR}{p_R}
\newcommand{\muR}{\mu_R}
\newcommand{\muRp}{\mu_R'}
\newcommand{\GamO}{\Gamma^{\Omega}}
\newcommand{\taua}{\tau_a}
\newcommand{\dr}{\delta\bm{r}}
\newcommand{\Vpec}{\bm{V}}
\newcommand{\uv}{\bm{u}}
\newcommand{\avg}[1]{\left\langle #1 \right\rangle}
\newcommand{\LOm}{L_{\Omega}}

\begin{document}

\title{Microscopic origin of the chiral pressure in a two-dimensional fluid}

\author{F. Vega Reyes}
\affiliation{Departamento de F\'isica, Universidad de Extremadura,
E-06006 Badajoz, Spain}
\affiliation{Instituto de Computaci\'on Cient\'ifica Avanzada (ICCAEx),
Universidad de Extremadura, E-06006 Badajoz, Spain}

\date{\today}

\begin{abstract}
  We show that a two-dimensional chiral fluid supports an antisymmetric contribution to the pressure, and
  determine its microscopic expression. Its origin lies outside the standard kinetic and collisional channels
  of the normal pressure tensor, stemming instead from the average of a parity-breaking pseudo-scalar. This
  chiral pressure is set by the rotational (vortex) viscosity of micropolar hydrodynamics, whose kinetic value
  for rough disks we compute here, to Gaussian order. Crucially, both the chiral pressure and the rotational
  viscosity survive at zeroth order in the gradients, meaning they persist even in the fully homogeneous fluid
  — which we show is not only well defined but steady (the \textit{Quiescent Chiral State}, QCS). Since this
  averaged pseudo-scalar does not vanish in a chiral fluid, any expansion of its distribution function must
  contain terms in this pseudo-scalar variable.
  % Finally, we
  % show that the chiral pressure in the QCS can only be sustained if the fluid retains a nonzero average spin:
  % the chiral fluid inherently spins in its most fundamental homogeneous steady state.
\end{abstract} 

\maketitle

\textit{Introduction}.- Chiral fluids have recently drawn great interest, due to their applications in
biological and artificial systems. As a consequence, the theory in this field has experienced a notable
progress: Avron~\cite{Avron1998} discussed the theoretical origin of the odd viscosity, that has its analog in
magnetized plasmas~\cite{Braginskii1965}, whereas the purely antisymmetric contributions to the stress tensor
and the complete form of the spin flux were discussed more
recently~\cite{Banerjee2017,Lou2022,Marconi2026,VegaReyes2026}. Kinetic theory of chiral fluids has also spun
a number of valuable works that try and apply the methods of the formal theory of non-homogeneous gases into
the realm of chiral fluids~\cite{Eren2026,Maire2026,Marconi2026,Maire2026b,Tsuzuki2026}.
 
However, these works miss one of the key distinguishing features that were uncovered for granular fluids with
passive rotations: Translational and rotational degrees of freedom are inherently correlated in an inelastic
hard rough particles fluid ~\cite{Brilliantov2007,VegaReyes2014}. As brilliantly stated by Brilliantov,
P\"oschel and Kranz~\cite{Brilliantov2007}, the force of that result lies not in the magnitude of the
correlation but in its being structurally unavoidable: it follows from the very collision operator already in
use, so that neglecting it is not a controlled approximation but an internal inconsistency.

It is thus hard to imagine that, in the same kind of fluid — inelastic hard rough particles, now with active
instead of passive rotations — this correlation would simply vanish. And yet, no attention has been devoted to
this crucial feature of fluids with particle rotation degrees of freedom. Moreover, the results by Santos
\cite{Santos2018} put into evidence that pseudoscalar moments are well defined even in the case of passive
rotations; it is just that their averages are inherently zero. Furthermore, Lopez-Casta\~no et
al.~\cite{LopezCastano2022} recently showed that a pseudoscalar moment of the velocity distribution function
denoted initially as \textit{bend coefficient} characterizes the velocity distribution of an experimental set
up of the chiral Stokes cavity~\cite{VegaReyes2026}, providing experimental evidence about the need of a
broken parity moment in the chiral fluid velocity distribution function. We build our results in this work
upon these three achievements (translations and rotations are inherently
correlated~\cite{Brilliantov2007,VegaReyes2026}, the existence of an antisymmetric contribution to the stress
tensor~\cite{Banerjee2017}, experimental evidence that a pseudoscalar moment with nonzero value enters into
the description of the distribution function of chiral kinetic theory \cite{LopezCastano2022}), showing that
in the two-dimensional fluid with active rotations the antisymmetric stress is generated by a pseudoscalar
moment of the particle distribution: the orbital moment, i.e.\ the angular momentum of the particles relative
to the local rotation of the flow. Since this moment does not vanish in general, and in particular not at
zeroth order in the gradients, any expansion of the distribution function of the chiral fluid must contain
terms in this pseudoscalar variable, a requirement that descriptions omitting the spin, or eliminating it
adiabatically, cannot meet. We demonstrate that the chiral pressure $p_R$ is proportional to this moment and
compute the proportionality coefficient exactly from the collision rule, providing a microscopic expression
for this purely antisymmetric contribution of the stress tensor in the chiral fluid.

Finally, we show that the natural homogeneous reference state of the chiral fluid, i.e.\ the zeroth-order
state of a perturbative solution, is not only well defined but steady: it is hydrostatic, completely
homogeneous, and has a constant nonzero average particle spin. We call it the \textit{quiescent chiral state}
(QCS, see companion work \cite{VegaReyes2026} for more detail); it should play for the kinetic theory of
chiral fluids the role that the homogeneous cooling state (HCS) plays for granular
fluids~\cite{Haff1983,Goldhirsch1993,vanNoije1998}. We compute the chiral pressure in the QCS from the general
microscopic expression.

\textit{The broken parity moment of the distribution function}.- In a physically chiral fluid, symmetry
breaking has a single origin --- the structural asymmetry of the particle --- expressed at  two levels:
at particle--fluid level, giving an external active torque $\taua$, and a particle--particle level. This differs
from models where parity is broken by the interaction law rather than by the matter~\cite{Maire2026,
  Marconi2026}, in which chirality resides in a prescribed tangential
impulse at contact. Existing kinetic treatments either omit the spin degree of freedom
altogether~\cite{Maire2026} or retain it microscopically but extract the response coefficients after adiabatic
elimination~\cite{Eren2026}. Both recover the algebraic form of the antisymmetric stress; as shown below,
however, the chiral pressure is built from the mismatch $\Omega-\omega/2$ ($\Omega$ is the angular velocity
average field and $\omega$ is the fluid vorticity), which requires the mean spin to be
retained as a hydrodynamic field.

The underlying reason is that so far kinetic-theory treatments of chiral and micropolar fluids have not
identified the existence of a parity-odd moment of the distribution function. By symmetry, such a moment is
the only possible microscopic source of an antisymmetric (chiral) contribution to the stress tensor: an
average built from a parity-even generating function cannot yield a pseudoscalar output.

The Hodge dual of the stress tensor carries, once the volumetric contribution is excluded, a mechanical
action of \emph{orbital} character~\cite{VegaReyes2026}. In a two-dimensional chiral fluid this
antisymmetric stress is captured by the chiral pressure~\cite{Shliomis1972,VegaReyes2026}
\begin{equation}
  \label{eq:chiral_pressure}
  p_R = \mu_R\!\left(\Omega - \frac{\omega}{2}\right),
\end{equation}
where $\Omega$ is the mean spin field, $\omega=\nabla\times\uv$ the vorticity, and
$\Omega=\omega/2$ is the co-rotation condition~\cite{Batchelor1967} at which $p_R$ vanishes.

We show now that such an action can only be generated by a transfer of orbital angular momentum within the
fluid element, of the form $\ell_O \equiv \langle \,\dr\times (m \Vpec) \rangle$, where $\dr$ is a structural
vector yielding a torque distance and $\Vpec$ is the particle velocity measured with respect to the local
fluid velocity. In the definition of $\ell_O$ and other places, please note that in 2D the vectorial product
of two vectors lies entirely in the Z direction, and hence its result can be considered as a (pseudo-)scalar,
and that is why, for instance, we write $\ell_O$ as scalar.

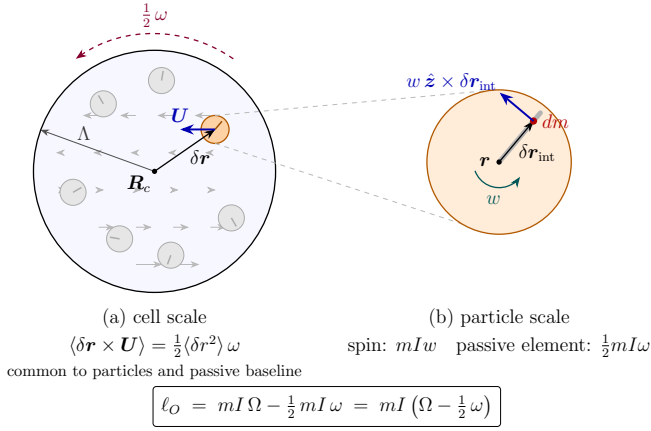
\begin{figure}
  \centering
  \resizebox{\columnwidth}{!}{%
\begin{tikzpicture}[>=Stealth, font=\large]
%% (a) cell scale
\def\Rc{2.6}
\fill[blue!3] (0,0) circle (\Rc);
\draw[thick] (0,0) circle (\Rc);
% background flow: simple shear (strain + rotation)
\foreach \x in {-2.0,-1.2,...,2.0}{
  \foreach \y in {-2.0,-1.2,...,2.0}{
    \pgfmathparse{(\x*\x+\y*\y<5.0) ? 1 : 0}
    \ifdim\pgfmathresult pt>0.5pt
      \draw[->,gray!55,line width=0.5pt] (\x,\y) -- ++({-0.28*\y},0);
    \fi
  }}
% particles, isotropically placed
\foreach \px/\py/\ang in {-1.1/1.45/120, -1.75/-0.45/30, 0.35/-1.8/250,
                          1.65/-1.0/300, 0.15/1.95/80, -0.75/-1.45/170}{
  \fill[gray!20] (\px,\py) circle (0.28);
  \draw[gray!65] (\px,\py) circle (0.28);
  \draw[gray!65,line width=0.9pt] (\px,\py) -- ++(\ang:0.21);
}
\coordinate (P) at (1.3,0.9);
\fill[orange!35] (P) circle (0.30);
\draw[orange!70!black,thick] (P) circle (0.30);
\draw[orange!60!black,line width=0.9pt] (P) -- ++(50:0.23);
% cell radius Lambda
\draw[->,black!70,thin] (0,0) -- (160:\Rc) node[pos=0.62,above=0pt,black!80] {$\Lambda$};
% cell centre and delta r
\fill (0,0) circle (1.6pt) node[below left=-1pt] {$\bm R_c$};
\draw[->,thick] (0,0) -- (P) node[pos=0.55,below right=-2pt] {$\delta\bm r$};
\draw[->,very thick,blue!70!black] (P) -- ++(-0.75,0) node[above=1pt] {$\bm U$};
% rigid-rotation part of the local flow
\draw[->,dashed,thick,purple!70!black] (55:2.95) arc (55:125:2.95);
\node[purple!70!black] at (90:3.35) {$\tfrac{1}{2}\,\omega$};
\node at (0,-3.15) {(a) cell scale};
\node at (0,-3.75) {$\langle\delta\bm r\times\bm U\rangle=\tfrac{1}{2}\langle\delta r^{2}\rangle\,\omega$};
\node[font=\normalsize] at (0,-4.3) {common to particles and passive baseline};
%% (b) particle scale
\coordinate (C) at (7.4,0.2);
\def\Rp{1.55}
\draw[gray!55,dashed,thin] ($(P)+(90:0.30)$)  -- ($(C)+(95:\Rp)$);
\draw[gray!55,dashed,thin] ($(P)+(270:0.30)$) -- ($(C)+(265:\Rp)$);
\fill[orange!15] (C) circle (\Rp);
\draw[orange!70!black,thick] (C) circle (\Rp);
\draw[gray!60,line width=3pt,line cap=round] (C) -- ($(C)+(50:1.38)$);
\coordinate (A) at ($(C)+(50:1.15)$);
\draw[->,thick] (C) -- (A) node[pos=0.5,below right=-2pt] {$\delta\bm r_{\rm int}$};
\fill[red!70!black] (A) circle (2.2pt);
\node[red!70!black,right=2pt] at (A) {$dm$};
\draw[->,very thick,blue!70!black] (A) -- ++(140:0.95)
      node[above left=-2pt] {$w\,\hat{\bm z}\times\delta\bm r_{\rm int}$};
\draw[->,thick,teal!70!black] ($(C)+(200:0.55)$) arc (200:325:0.55);
\node[teal!70!black] at ($(C)+(262:0.82)$) {$w$};
\fill (C) circle (1.6pt) node[left=2pt] {$\bm r$};
\node at (7.4,-3.15) {(b) particle scale};
\node at (7.4,-3.75) {spin: $mIw$\quad passive element: $\tfrac{1}{2}mI\omega$};
%% assembly
\node[draw,rounded corners=2pt,inner sep=5pt,fill=white] at (3.7,-5.05)
  {$\ell_O \;=\; mI\,\Omega-\tfrac{1}{2}\,mI\,\omega \;=\; mI\left(\Omega-\tfrac{1}{2}\,\omega\right)$};
\end{tikzpicture}}
  \caption{Hydrodynamic derivation of the orbital moment. (a) Coarse-graining cell of radius $\Lambda$
    centred at $\bm R_c$, with particles distributed isotropically in a background flow that contains both
    strain and rotation. The cell-scale orbital term, Eq.~\eqref{eq:orbital-vorticity}, is common to the
    particles and to the passive baseline and cancels. (b) Particle scale: a mass element $dm$ at
    $\delta\bm r_{\rm int}$ from the particle centre contributes $mIw$ through the spin, against
    $\tfrac12 mI\omega$ for a passive fluid element with the same mass distribution. The difference gives
    Eq.~\eqref{eq:ell-avg}.}
  \label{fig:hydro}
\end{figure}

\textit{The orbital angular momentum derived in a hydrodynamic unit cell}.- We derive $\ell_O$
hydrodynamically (Fig.~\ref{fig:hydro}). The idea is to compare the angular momentum of the
particles in a coarse-graining cell with that of the \emph{passive} fluid elements that would occupy their
place, i.e.\ elements with the same mass distribution moving with the local flow $\uv$. Only this difference
can generate an antisymmetric stress: the absolute angular momentum also changes under a superposed rigid
rotation, $w\to w+\Omega_f$, $\omega\to\omega+2\Omega_f$, which produces no internal stress. Let $\bm R_c$ be
the center of a cell of radius $\Lambda$ and
\begin{equation}
  \label{eq:deltar}
\dr=\bm r-\bm R_c,
\qquad
\uv_c=\uv(\bm R_c),
\end{equation}
with $\bm r$ the center of a particle in the cell, and let $\dr_{\rm int}$ be the position of a mass element
$dm$ relative to that center, so that $\int dm\,\dr_{\rm int}=0$. By K\"onig's theorem, the angular momentum
about $\bm R_c$ of a particle, and that of its passive counterpart, each splits into an orbital part, carried
by the center at $\dr$, and an internal part about that center. The comparison thus involves four terms, which
we evaluate in turn.

\emph{(i) Particle, orbital part.} The particle center moves with $\dot{\bm r}=\uv(\bm r)+\bm c$, where the
peculiar velocity $\bm c$ has zero local average by definition of $\uv$. Hence $\avg{\dr\times\bm c}=0$ and,
since $\avg{\dr}\times\uv_c=0$ as well, this term averages to $m\avg{\dr\times\bm U}$, with
\begin{equation}
  \label{eq:V_expansion}
  \bm U \equiv \uv(\bm r)-\uv_c \simeq(\dr\cdot\nabla)\uv\big|_{\bm R_c}.
\end{equation}
Averaging over the cell,
\begin{equation}
  \label{eq:avg_delta_times_V}
\left\langle
\dr\times \bm U \right\rangle=
\epsilon_{ik}\left\langle \delta r_i\delta r_j \right\rangle_c\partial_j u_k =
\frac{1}{2}\left\langle\delta r^2\right\rangle\epsilon_{ik}\,\partial_i u_k ,
\end{equation}
where we used that, for particles distributed isotropically in the cell,
$\left\langle\delta r_i\delta r_j\right\rangle_c=\frac{1}{2}\left\langle\delta r^2\right\rangle\delta_{ij}$
in two dimensions. With the 2D vorticity
$\omega=(\nabla\times\uv)_z=\epsilon_{ik}\partial_iu_k=\partial_xu_y-\partial_yu_x$,
\begin{equation}
\left\langle
\dr\times\bm U
\right\rangle = \frac{1}{2}\left\langle\delta r^2\right\rangle\,\omega ,
\label{eq:orbital-vorticity}
\end{equation}
positive for $\omega>0$, as for a rigid rotation $\uv=\tfrac{1}{2}\omega\,\hat{\bm z}\times\bm r$. Only the
antisymmetric part of $\nabla\uv$ survives the isotropic average: the strain drops out, and what remains is
the local rotation of the fluid, with angular velocity $\omega/2$. The term grows with the cell size,
$\avg{\delta r^2}=\Lambda^2/2$ for a uniform distribution.

\emph{(ii) Particle, internal part.} The particle rotates rigidly with angular velocity $w$, so that
\begin{equation}
  \int dm\;\dr_{\rm int}\times\left(w\,\hat{\bm z}\times\dr_{\rm int}\right)
  = w\int dm\,|\dr_{\rm int}|^{2} = mI\,w ,
  \label{eq:spin-contribution}
\end{equation}
where $I$ is the micromoment of inertia~\cite{Eringen1966,VegaReyes2026}, i.e.\ the moment of inertia per
unit mass (the squared radius of gyration), $I=\kappa\sigma^2/4$ with $\kappa$ the reduced moment of inertia.
This term averages to $mI\Omega$, with $\Omega\equiv\langle w\rangle$.

\emph{(iii) Passive element, orbital part.} Its center moves exactly with $\uv(\bm r)$, with no peculiar
velocity. This term is therefore $m\avg{\dr\times\uv(\bm r)}=m\avg{\dr\times\bm U}$, identical to (i) term by
term, before any gradient expansion.

\emph{(iv) Passive element, internal part.} Each of its mass elements moves with the flow at its own position,
\begin{align}
  & \int dm\;\dr_{\rm int}\times\uv(\bm r+\dr_{\rm int})
  \; \\
  & \simeq\;\epsilon_{ik}\,\partial_j u_k\int dm\;\delta r_{{\rm int},i}\,\delta r_{{\rm int},j}
  \;=\;\tfrac{1}{2}\,mI\,\omega ,
  \label{eq:flow-internal}
\end{align}
with $\int dm\,\delta r_{{\rm int},i}\,\delta r_{{\rm int},j}=\tfrac{1}{2}mI\,\delta_{ij}$ for the isotropic mass
distribution of a disk. This is the contraction of Eq.~\eqref{eq:avg_delta_times_V}, now at the particle
scale: a body of moment of inertia $mI$ co-rotating with the fluid at $\omega/2$.

\emph{The difference.} Subtracting (iii) from (i), the cell-scale term cancels identically, and with it all
dependence on $\Lambda$. Subtracting (iv) from (ii), the orbital moment per particle is
\begin{equation}
  \ell_O \;=\; mI\,\Omega \;-\; \tfrac{1}{2}\,mI\,\omega
  \;=\; mI\left(\Omega-\frac{\omega}{2}\right),
  \label{eq:ell-avg}
\end{equation}
The same geometry thus acts at two scales: at the cell scale
it affects particles and passive elements alike and cancels; at the particle scale it compares the actual
spin $w$ with the local rotation of the fluid, $\omega/2$. Neither coefficient is postulated: $\Omega$ and
$\tfrac12\omega$ share the weight $mI$ because Eqs.~\eqref{eq:spin-contribution} and \eqref{eq:flow-internal}
are integrals over the \emph{same} mass distribution, and the $\tfrac12$ is the tensor contraction of
Eq.~\eqref{eq:flow-internal}. Corrections from the curvature of the flow are $O(mI^{2}\nabla^{2}\omega)$, of
second order in gradients.

\textit{Orbital angular momentum from the microscopic kinematics of particles}.-

\begin{figure}
  \centering
\begin{tikzpicture}[>=Stealth, font=\small]
% ---------- background flow u(r): simple shear, u_x = -gamma y ----------
\foreach \y in {-2.3,-1.5,-0.7,0.7,1.5,2.3}{
  \foreach \x in {-3.6,-2.7,-1.8,-0.9,0,0.9,1.8,2.7,3.6}{
    \pgfmathsetmacro{\dc}{sqrt(\x*\x+\y*\y)}
    \pgfmathsetmacro{\de}{sqrt((\x+3.0)*(\x+3.0)+(\y-1.85)*(\y-1.85))}
    \pgfmathsetmacro{\dl}{sqrt((\x-2.9)*(\x-2.9)+(\y+2.75)*(\y+2.75))}
    \pgfmathparse{(\dc>2.5 && \de>1.15 && \dl>0.6) ? 1 : 0}
    \ifdim\pgfmathresult pt>0.5pt
      \draw[->,gray!60,line width=0.6pt] ({\x+0.15*\y},\y) -- ({\x-0.15*\y},\y);
    \fi
  }}
\node[gray!50!black,anchor=east] at (4.0,-2.95) {local background flow $\bm u(\bm r)$};
% ---------- local fluid rotation: vorticity ----------
\coordinate (E) at (-3.0,1.85);
\draw[purple!70!black,dashed] (E) circle (0.32);
\draw[->,thick,purple!70!black] ($(E)+(-30:0.52)$) arc (-30:250:0.52);
\node[purple!70!black,anchor=west] at ($(E)+(0.58,0.12)$) {$\tfrac12\omega$};
\node[purple!70!black] at ($(E)+(0,-0.85)$) {$\omega=\nabla\times\bm u$};
% ---------- the particle, rigidly rotating with w ----------
\def\Rp{1.5}
\def\b{1.1}
\fill[orange!15] (0,0) circle (\Rp);
\draw[orange!70!black,thick] (0,0) circle (\Rp);
\draw[gray!70,dashed] (0,0) circle (\b);
\node[gray!50!black,font=\footnotesize] at (0,-0.72) {orbit $|\delta\bm r|=b$};
% spin of the whole particle
\draw[->,very thick,teal!70!black] (200:1.78) arc (200:335:1.78);
\node[teal!70!black] at (268:2.12) {$w$};
% centre and active point
\coordinate (A) at (35:\b);
\fill (0,0) circle (1.6pt);
\node[anchor=north east] at (-0.03,-0.03) {$\bm r_c$};
\draw[->,thick] (0,0) -- (A) node[pos=0.5,anchor=north west,inner sep=1.5pt] {$\delta\bm r$};
% velocities at the active point
\draw[->,very thick,teal!70!black] (A) -- ++(125:1.0)
   node[anchor=south east,inner sep=1pt] {$w\,\hat{\bm z}\times\delta\bm r$};
\draw[->,very thick,blue!70!black] (A) -- ++(55:1.05) node[anchor=west,inner sep=2pt] {$\bm V$};
\fill[red!70!black] (A) circle (2.2pt);
\node[red!70!black,anchor=west] at (1.52,0.5) {$\bm r_a$};
\end{tikzpicture}
  \caption{Microscopic derivation of the orbital moment. The disk rotates rigidly with angular velocity $w$
    (teal), carrying the active point $\bm r_a$, at $\dr$ from its center $\bm r_c$, along the orbit $|\dr|=b$.
    It is immersed in a local background flow $\uv(\bm r)$ (gray), whose vorticity $\omega=\nabla\times\uv$ is a
    local rotation of the fluid at $\omega/2$ (purple). The peculiar velocity $\Vpec$ (blue) is measured
    relative to the flow at $\bm r_a$. Averaged over orientations, the spin contributes $b^2\Omega$ to
    $\dr\times\Vpec$, and the circulation of $\uv$ around the orbit $\tfrac12 b^2\omega$,
    Eq.~\eqref{eq:L-exact}.}
  \label{fig:active-point}
\end{figure}
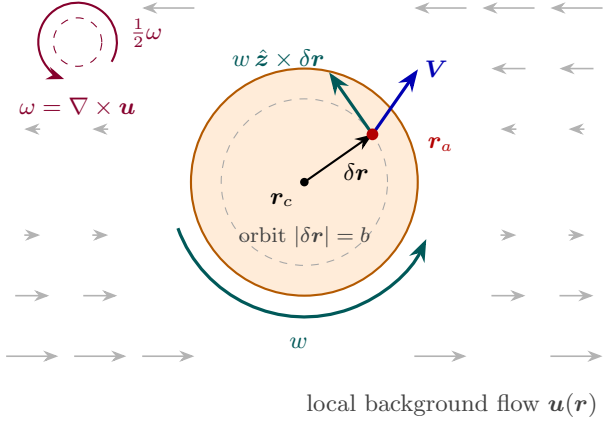

We now define the structural vector $\dr$ as the one from a
particle's own center $\bm r_c$ to its marked active point~\cite{LopezCastano2022}, of fixed length
$b\le\sigma/2$, and $\Vpec\equiv\dot{\bm r}_a-\uv(\bm r_a)$ as the peculiar velocity of that active point,
i.e.\ its velocity relative to the local flow evaluated \emph{at the active point itself}. Rigidity of the
particle gives $\dot{\dr}=w\,\hat{\bm z}\times\dr$, with $w$ the particle's own angular velocity, so that
\begin{equation}
  \Vpec \;=\; \bm c \;+\; w\,\hat{\bm z}\times\dr \;-\;\left[\uv(\bm r_c+\dr)-\uv(\bm r_c)\right],
  \label{eq:rigid-split}
\end{equation}
where $\bm c\equiv\dot{\bm r}_c-\uv(\bm r_c)$ is the ordinary peculiar velocity of the particle's own
center (see Fig.~\ref{fig:active-point}).

Using the 2D identity $\dr\times(\hat{\bm z}\times\dr)=|\dr|^{2}=b^2$,
\begin{equation}
  \dr\times\Vpec \;=\; \dr\times\bm c \;+\; \dr\times\uv(\bm r_c) \;+\; w\,b^{2}
  \;-\; \dr\times\uv(\bm r_c+\dr).
  \label{eq:drxV-split}
\end{equation}
The orientation angle $\theta$ of $\dr$ is a cyclic variable: the collision rule of rough disks does not
depend on it, so that $\theta(t)=\theta(0)+\int_0^t w\,dt'$ with a trajectory that is independent of
$\theta(0)$. For uniformly distributed initial orientations, $\theta$ therefore remains uniformly distributed
and statistically independent of the position, velocity and spin of the particle at all times. The first two
terms of Eq.~\eqref{eq:drxV-split} thus average \emph{exactly} to zero, and the last one is fixed by Stokes'
theorem on the circle $|\dr|=b$,
\begin{equation}
  \avg{\dr\times\uv(\bm r_c+\dr)}_\theta \;=\; \frac{1}{2\pi}\oint_{|\dr|=b}\uv\cdot d\bm l
  \;=\; \tfrac{1}{2}\,b^{2}\,\bar\omega_b ,
  \label{eq:stokes}
\end{equation}
where $\bar\omega_b=\omega+\tfrac{1}{8}b^{2}\nabla^{2}\omega+\dots$ is the vorticity averaged over the disk of
radius $b$ centered at $\bm r_c$. Averaging over the particles in the coarse-graining volume, with
$\Omega\equiv\avg{w}$, gives
\begin{align}
   m \avg{\dr\times\Vpec} \;= &\; m\, b^{2}\left(\Omega-\tfrac{1}{2}\,\bar\omega_b\right)
  \;= \nonumber \\
  & m\, b^{2}\left(\Omega-\tfrac{1}{2}\,\omega\right)+O\!\left(mb^{4}\nabla^{2}\omega\right),
  \label{eq:L-exact}
\end{align}
with no truncation involved: the correction is of second order in gradients and does not enter at the
orders considered here. Both the relative sign and the factor $1/2$ follow from the geometry rather than
being postulated: $\Vpec$ is referred to the flow \emph{at the active point}, while $\bm c$ is referred to the
flow \emph{at the particle's own center}, and the factor $1/2$ is the circulation of the flow around the orbit
of the active point. Since spin and flow contributions involve the \emph{same} vector $\dr$, their common
weight $b^{2}$ is an identity, not an approximation. The choice of active point only affects this weight:
every material point of the rigid disk is an admissible active point, and summing Eq.~\eqref{eq:L-exact}
over the mass distribution replaces $mb^2$ by $\int dm\,b^2=mI$, so that the orbital moment per particle is 
\begin{equation}
  \ell_O\;\equiv\;\int dm\,\avg{\dr\times\Vpec} \;=\; mI\left(\Omega-\tfrac{1}{2}\,\omega\right)
  +O\!\left(mI^{2}\nabla^{2}\omega\right),
  \label{eq:L-sum}
\end{equation}
in exact agreement with Eq.~\eqref{eq:ell-avg}. The length $b$, of order the particle radius, is consistent
with the lower bound of the averaging window.

\textit{The chiral pressure}.- We have shown by two independent routes that the orbital moment per
particle is $\ell_O=mI(\Omega-\omega/2)$: the combination $\Omega-\omega/2$ does not depend on the reference
point (the cell scale or any material point of the particle) up to corrections of second order in gradients,
and its physical normalization is the moment of inertia $mI$. We now show that the chiral pressure is the
mechanical response of the fluid to the orbital angular-momentum density
$\mathcal{L}_O\equiv n\ell_O=\rho I(\Omega-\omega/2)$, with $\rho=nm$ the mass density,
\begin{equation}
  p_R  =  \chi_R\,\mathcal{L}_O ,
  \label{eq:chiral-susceptibility}
\end{equation}
with a susceptibility $\chi_R$ that is not postulated but determined below from the collision rule;
comparison with Eq.~\eqref{eq:chiral_pressure} then gives $\mu_R=\rho I\chi_R$.

The sign in Eq.~\eqref{eq:ell-avg} is physically meaningful: $\tfrac12 mI\omega$ in
Eq.~\eqref{eq:flow-internal} is the angular momentum that a passive fluid element with the same mass
distribution would carry due to the background flow alone. When $\ell_O > 0$, the particles spin faster than this passive baseline,
generating a net excess of angular momentum — the chiral mismatch — which is precisely what $p_R$ measures.
The co-rotation condition $\Omega = \omega/2$ sets $\ell_O = 0$ and hence $p_R = 0$: no mismatch,
no chiral pressure.

\textit{The averaging window}.-
The hydrodynamic derivation of Eq.~\eqref{eq:ell-avg} is defined relative to an averaging cell bounded on both
sides. It defines the averaging window

\begin{equation}
  \sigma/2 \;\lesssim\; \Lambda \;\ll\; \LOm \equiv |\Omega/\nabla\Omega| ,
  \label{eq:window}
\end{equation}
with $\Lambda$ the radius of the coarse-graining cell. The lower bound is the particle radius $\sigma/2$,
since $\dr\times\Vpec$ is intrinsically defined over a region of the size of the particle; the upper bound is
the scale at which the hydrodynamic variation of $\Omega$ would be filtered out. Condition~\eqref{eq:window}
is thus the condition of existence of $\Omega$ as a hydrodynamic field, and it closes exactly where the
continuum description ceases to resolve the spin. Within the window, $\ell_O$ does not depend on $\Lambda$:
the only $\Lambda$-dependent contribution, Eq.~\eqref{eq:orbital-vorticity}, is common to the particles and to
the passive baseline.

\textit{Collisional origin}.-
Unlike every transport coefficient of a simple fluid, $\muR$ enters at zeroth order in gradients: what it
multiplies, $\Omega$, is not the gradient of anything but a hydrodynamic field in its own right. It is
therefore computable from the homogeneous base state, without solving the first-order Chapman--Enskog integral
equation.  The collisional rate of change of the internal angular momentum of a rough-disk gas is known in
closed form~\cite{Santos2018} and is the exact kinetic counterpart of the source term $-2\pR$ in the spin
balance. It is entirely general, present for any rough gas, even with neither chirality nor
activity. 
In the homogeneous cooling state it vanishes identically because isotropy sets the mean angular velocity to
zero: the coefficient is there and the integral is computed, but the moment multiplying it is null. What
follows activates that term with $\Omega\neq0$, sustained by $\taua$ against $\GamO$.

\begin{figure}
  \centering
  \includegraphics[width=\columnwidth]{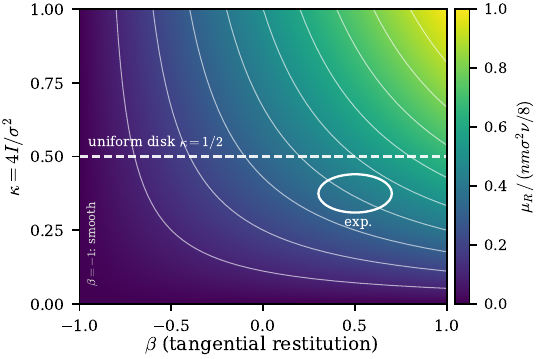}
  \caption{Rotational viscosity of rough disks at Gaussian order, Eq.~\eqref{eq:muR-exact}, normalized by
    $nm\sigma^2\nu/8$, in the plane of tangential restitution $\beta$ and reduced moment of inertia $\kappa$.
    It does not depend on the normal restitution coefficient. Dashed line: uniform disk, $\kappa=1/2$.
    Ellipse: approximate range of the experimental disks in a previous work \cite{LopezCastano2022}.}
  \label{fig:muR}
\end{figure}

The spin balance reads \cite{VegaReyes2026}
\begin{equation}
  \rho I\,\frac{D\Omega}{Dt} \;=\; \nabla\!\cdot\bm C \;+\; \taua \;-\; \GamO\,\Omega \;-\; 2\pR ,
  \label{eq:spin-balance}
\end{equation}
where $\taua=n\tau_p$ is the active torque density ($\tau_p$ the active torque on a single particle),
$\GamO=n\gamma_R$ the rotational drag density ($\gamma_R$ the rotational drag coefficient of a single
particle), and $\bm C$ the spin flux (couple stress)~\cite{VegaReyes2026}.
The collisional source $-2\pR$ follows from the collision rule of rough disks. For two disks at contact, let
$\hat{\bm\sigma}=(\bm r_2-\bm r_1)/\sigma$ and $\hat{\bm\sigma}_\perp=\hat{\bm z}\times\hat{\bm\sigma}$. The
tangential component of the relative velocity of the contact points is
$g_t=[\bm v_1-\bm v_2+\tfrac12\sigma(w_1+w_2)\hat{\bm\sigma}_\perp]\cdot\hat{\bm\sigma}_\perp$. Sampling the
flow over the contact separation $\bm r_1-\bm r_2=-\sigma\hat{\bm\sigma}$ and averaging over $\hat{\bm\sigma}$
at Gaussian order (the collision kernel $\Theta(\bm g\cdot\hat{\bm\sigma})\,\bm g\cdot\hat{\bm\sigma}$
factorizes from $g_t$),
\begin{equation}
  \langle g_t\rangle \;=\; \sigma\left(\Omega-\tfrac{1}{2}\,\omega\right) \;=\; \sigma\,\frac{\ell_O}{mI} .
  \label{eq:gt}
\end{equation}
The tangential contact velocity, the only quantity a rough collision responds to, thus measures exactly the
orbital moment: a third, independent appearance of $\Omega-\omega/2$, now at the contact scale. In each
collision the tangential impulse changes the spin of each partner by
$\Delta w=-\frac{1+\beta}{1+\kappa}\,g_t/\sigma$~\cite{Santos2018,VegaReyes2014}, so that the collisional
rate of change of spin density is $\rho I\nu\langle\Delta w\rangle=-\nu\frac{1+\beta}{1+\kappa}\,\mathcal{L}_O$.
Identifying it with $-2\pR$ in Eq.~\eqref{eq:spin-balance} yields Eq.~\eqref{eq:chiral-susceptibility} with
\begin{equation}
  \chi_R \;=\; \frac{1}{2}\,\nu\,\frac{1+\beta}{1+\kappa},
  \label{eq:chiR-exact}
\end{equation}
where $\nu=2n\chi\sigma\langle g\rangle$ is the collision frequency ($\chi$ the pair correlation at contact,
$\langle g\rangle$ the mean relative speed), $\beta\in[-1,1]$ the tangential restitution coefficient
($\beta=-1$ for smooth disks), and $\kappa=4I/\sigma^2$ the reduced moment of inertia. The susceptibility is
simply a rate, the frequency at which collisions exchange tangential momentum, and the chiral pressure is the
relative orbital angular-momentum density times the rate at which collisions transfer it. Equivalently, for
rough inelastic disks of diameter $\sigma$, mass $m$, and micromoment of inertia $I=\frac{1}{4}\kappa\sigma^2$,
the rotational viscosity is
\begin{equation}
  \muR \;=\; \rho I\chi_R \;=\; \frac{1}{2}\,\rho I\nu\,\frac{1+\beta}{1+\kappa} .
  \label{eq:muR-exact}
\end{equation}
Three properties follow immediately (Fig.~\ref{fig:muR}). First, $\muR>0$ at Gaussian order for any $\beta>-1$
and $\kappa>0$: at this order the rotational viscosity is sign-definite. Like any transport coefficient,
$\muR$ is a functional of the interparticle interaction, and for rough hard disks at Gaussian order the
negative branch $\muR<0$ does not exist. Since then also the modified rotational viscosity
$\muRp=\muR\GamO/[2(2\muR+\GamO)]$ is positive, the active branch of confined chiral flows, which requires
$\mu+\muRp<0$~\cite{VegaReyes2026}, is not accessible to this interaction at this order: only the dissipative,
screened branch is realized.  Whether an effective $\muR<0$ can arise from non-Gaussian corrections, which in
the QCS include the translational--rotational correlations, or requires a different interaction model, remains
open. Second, $\muR$ is parity-even — it depends only on $\beta$, $\kappa$, $n$ and $\nu$, none of which know
about chirality: at Gaussian order the same coefficient appears in any rough granular gas, active or
not. Third, $\muR\propto n\nu\propto n^2$: it is a collisional coefficient, one density order higher than the
kinetic shear viscosity $\mu\propto n^0$.

\textit{Assembly}.-
Combining the kinematic identities~\eqref{eq:ell-avg} and~\eqref{eq:L-sum} with the collisional
coefficient,
\begin{equation}
  \pR \;=\; \muR\!\left(\Omega-\tfrac{1}{2}\,\omega\right)
       \;=\; \chi_R\,\mathcal{L}_O .
  \label{eq:pR-assembled}
\end{equation}
This is not the output of a single calculation. Its ingredients are of different natures: $\mathcal{L}_O$ is
purely kinematic, an identity valid for an isolated particle and requiring no collisions whatsoever, while
$\muR$ is purely collisional. The chiral pressure has a twofold physical origin --- what the spin is, and how
it couples --- and cannot be closed by attending to the collisional part alone. A model without internal
structure has no $\dr$, so that $\avg{\dr\times\Vpec}$ vanishes identically and the antisymmetric stress it
reports has no microscopic support.

\textit{The quiescent chiral state}.-
In a homogeneous state at rest, $\omega=0$ and $\nabla\Omega=0$, and the stationary spin
balance~\eqref{eq:spin-balance} with Eq.~\eqref{eq:pR-assembled} gives
\begin{equation}
  \left(2\muR+\GamO\right)\Omega = \taua ,
  \qquad
  \pR = \psi\,\taua ,
  \qquad
  \psi \equiv \frac{\muR}{2\muR+\GamO} .
  \label{eq:QCS}
\end{equation}
In the QCS $\omega=0$: the cell-scale vortical contribution, Eq.~\eqref{eq:orbital-vorticity}, vanishes, and
the orbital moment reduces to $\ell_O=mI\Omega$, carried entirely by the spin, so that
$\pR=\chi_R\,\rho I\Omega$.
Activity injects angular momentum continuously, and that momentum must go somewhere. With the spin retained
there are two reservoirs: the internal spin, drained against $\GamO$, and the orbital motion; at rest the
former absorbs all of it and the state~\eqref{eq:QCS} exists, with the chiral pressure set by the entrainment
coefficient $\psi$. With the spin removed only the orbital reservoir remains, and the injected momentum has
nowhere to go but the flow.

\textit{Conclusions}.- We have shown that the chiral pressure of a two-dimensional fluid of rough, actively
rotating particles has a twofold microscopic origin. Its kinematic part is the orbital moment
$\ell_O=\avg{\dr\times(m\Vpec)}=mI(\Omega-\omega/2)$, the angular momentum of the particles relative to that of
a passive element of the flow with the same mass distribution. We obtained it by two independent routes --- a
coarse-graining cell, where the cell-scale orbital term cancels against the passive baseline, and the active
point of each particle, where it follows from Stokes' theorem --- and it reappears at the contact scale as the
mean tangential contact velocity, $\langle g_t\rangle=\sigma\ell_O/(mI)$. In all three cases the factor $1/2$ and
the common weight $mI$ are exact consequences of geometry. Its collisional part is the rotational viscosity,
which we computed as $\muR=\tfrac12 \rho I\nu(1+\beta)/(1+\kappa)$ at Gaussian order for inelastic rough disks.
Because it multiplies a hydrodynamic field rather than a gradient, $\muR$ is the only viscosity that enters at
zeroth order, and it follows from the homogeneous base state without solving the Chapman--Enskog integral
equation. It is positive for any $\beta>-1$, so that for hard rough disks at this order only the screened
branch of confined chiral flows is accessible~\cite{VegaReyes2026}.

Since both parts survive at zeroth order, the chiral pressure persists in the homogeneous steady state at
rest, the quiescent chiral state (QCS \cite{VegaReyes2026}), where $\pR=\psi\taua$ is carried entirely by the spin.

The nonvanishing of $\ell_O$ at zeroth order has a direct consequence for kinetic theory: the reference
distribution function of a chiral fluid cannot be parity-even, and must depend on the pseudoscalar
$\dr\times\Vpec$. A precursor of this dependence, the bend coefficient, was already identified in the
experimental velocity statistics of chiral disks and its accompanying theoretical anaylsis, where a
distribution function with a pseudo-scalar moment was proposed~\cite{LopezCastano2022}. The variable
$\dr\times\Vpec$ itself mixes translational and rotational degrees of freedom: its mean involves only
one-particle marginals, whereas its fluctuations --- those of the tangential contact velocity $g_t$ --- are
precisely the channel through which the collision rule generates translational--rotational correlations. Such
a dependence is also the natural place for the translational--rotational correlations that are structurally
unavoidable in rough-particle fluids~\cite{Brilliantov2007,VegaReyes2014}. Our Gaussian-order results do not
yet include them; they enter through the non-Gaussian corrections to $\muR$, where the possibility of a sign
change remains open. The complete distribution function of the quiescent chiral state is the subject of work
in progress~\cite{VegaReyes2026_f}.

\begin{acknowledgments}
  I acknowledge support from Ministerio de Ciencia, Innovaci\'on y Universidades (AEI) and ERDF ``A way of
  making Europe'' through project no. MCIN/AEI/10.13039/501100011033 and fellowship PRX21/00490. I also
  acknowledge support from Junta de Extremadura through contract No. GR24077.
\end{acknowledgments}

%% ============================================================

\bibliography{chiral_pressure}

\end{document}